\documentclass[a4paper,10pt,twoside]{cpc-hepnp}
\usepackage{CJK,upgreek,fancyhdr}
\usepackage{multicol}
\usepackage{graphicx}
\usepackage{booktabs}
\usepackage{amssymb,bm,mathrsfs,bbm,amscd}
\usepackage[tbtags]{amsmath}
\usepackage{lastpage}

\usepackage[english]{babel}
\usepackage[colorlinks=false,pdfpagemode=FullScreen,,setpagesize=off,pdfborder={0 0 0}]{hyperref}
\usepackage{overpic}
\usepackage{lineno}
\usepackage{color}

\begin{document}
\begin{CJK*}{UTF8}{gkai}
%% note here we use UTF8 and gbsn(simpe chinese character)
%% however, in author_list_en_cn.tex, there is a traditional chinese character for the name
%% S.~L.~Olsen(馬鵬)
%% so in there, we use the {\CJKfamily{bsmi}é¦¬é µ¬} to show that name
%% and for the name 郭玥,the'玥'can not be shown in gbsn,
%% and for the name 康晓珅, {\CJKfamily{bsmi}珅}
%% so we also use {\CJKfamily{bsmi}玥} to show it
%% If editors want to change the font, please notice that problem.

\fancyhead[c]{\small Chinese Physics C~~~Vol. xx, No. x (201x) xxxxxx}
\fancyfoot[C]{\small 010201-\thepage}

\footnotetext[0]{Received xxxx June xxxx}

\title{Measurement of integrated luminosity and center-of-mass energy of data taken by BESIII at
$\mathbf{\sqrt{s} = 2.125}$ GeV\thanks{
Supported in part by National Key Basic Research Program of China under Contract No. 2015CB856700; National Natural Science Foundation of China (NSFC) under Contracts Nos. 11235011, 11322544, 11335008, 11425524, 11635010, 11675184, 11735014; the Chinese Academy of Sciences (CAS) Large-Scale Scientific Facility Program; the CAS Center for Excellence in Particle Physics (CCEPP); the Collaborative Innovation Center for Particles and Interactions (CICPI); Joint Large-Scale Scientific Facility Funds of the NSFC and CAS under Contracts Nos. U1232201, U1332201, U1532257, U1532258; CAS under Contracts Nos. KJCX2-YW-N29, KJCX2-YW-N45; 100 Talents Program of CAS; National 1000 Talents Program of China; INPAC and Shanghai Key Laboratory for Particle Physics and Cosmology; German Research Foundation DFG under Contracts Nos. Collaborative Research Center CRC 1044, FOR 2359; Istituto Nazionale di Fisica Nucleare, Italy; Koninklijke Nederlandse Akademie van Wetenschappen (KNAW) under Contract No. 530-4CDP03; Ministry of Development of Turkey under Contract No. DPT2006K-120470; National Natural Science Foundation of China (NSFC) under Contract No. 11505010; The Swedish Resarch Council; U. S. Department of Energy under Contracts Nos. DE-FG02-05ER41374, DE-SC-0010118, DE-SC-0010504, DE-SC-0012069; U.S. National Science Foundation; University of Groningen (RuG) and the Helmholtzzentrum fuer Schwerionenforschung GmbH (GSI), Darmstadt; WCU Program of National Research Foundation of Korea under Contract No. R32-2008-000-10155-0.
}}

\maketitle
\begin{small}
\begin{center}
M.~Ablikim(麦迪娜)$^{1}$, M.~N.~Achasov$^{9,e}$, S. ~Ahmed$^{14}$, X.~C.~Ai(艾小聪)$^{1}$, O.~Albayrak$^{5}$, M.~Albrecht$^{4}$, D.~J.~Ambrose$^{44}$, A.~Amoroso$^{49A,49C}$, F.~F.~An(安芬芬)$^{1}$, Q.~An(安琪)$^{46,a}$
% ...
\\
\vspace{0.2cm}
(BESIII Collaboration)\\
\vspace{0.2cm} {\it
$^{1}$ Institute of High Energy Physics, Beijing 100049, People's Republic of China\\
% ...
$^{4}$ Bochum Ruhr-University, D-44780 Bochum, Germany\\
$^{5}$ Carnegie Mellon University, Pittsburgh, Pennsylvania 15213, USA\\
% ...
$^{9}$ G.I. Budker Institute of Nuclear Physics SB RAS (BINP), Novosibirsk 630090, Russia\\
% ...
$^{14}$ Helmholtz Institute Mainz, Johann-Joachim-Becher-Weg 45, D-55099 Mainz, Germany\\
% ...
$^{44}$ University of Rochester, Rochester, New York 14627, USA\\
% ...
$^{46}$ University of Science and Technology of China, Hefei 230026, People's Republic of China\\
% ...
$^{49}$ (A)University of Turin, I-10125, Turin, Italy; (B)University of Eastern Piedmont, I-15121, Alessandria, Italy; (C)INFN, I-10125, Turin, Italy\\
% ...
$^{a}$ Also at State Key Laboratory of Particle Detection and Electronics, Beijing 100049, Hefei 230026, People's Republic of China\\
% ...
$^{e}$ Also at the Novosibirsk State University, Novosibirsk, 630090, Russia\\
% ...
}\end{center}

\vspace{0.4cm}
\end{small}

%%\linenumbers

\begin{abstract}
  To study the nature of the state $Y(2175)$, a dedicated data set of
  $e^+e^-$ collision data was collected at the center-of-mass energy of 2.125 GeV  with the
  BESIII detector at the BEPCII collider. By analyzing large-angle Bhabha
  scattering events, the integrated luminosity of this data set is
  determined to be $108.49\pm0.02\pm0.85$~pb$^{-1}$, where the first
  uncertainty is statistical and the second one is systematic. In
  addition, the center-of-mass energy of the data set is determined
  with radiative dimuon events to be $2126.55\pm0.03\pm0.85$~MeV, where the
  first uncertainty is statistical and the second one is systematic.
\end{abstract}

\begin{keyword}
Bhabha scattering, luminosity, radiative dimuon events, center-of-mass energy
\end{keyword}

\begin{pacs}
13.66.De, 13.66.Jn
\end{pacs}

%\footnotetext[0]{\hspace*{-3mm}\raisebox{0.3ex}{$\scriptstyle\copyright$}2013
%Chinese Physical Society and the Institute of High Energy Physics
%of the Chinese Academy of Sciences and the Institute
%of Modern Physics of the Chinese Academy of Sciences and IOP Publishing Ltd}%

\begin{multicols}{2}

\section{Introduction}

The state $Y(2175)$, denoted as $\phi(2170)$ in Ref.~\cite{pdg}, was first
observed by the BaBar experiment~\cite{babar_y2175_1, babar_y2175_2}
in the initial-state-radiation (ISR) process
$e^{+}e^{-}\rightarrow\gamma_{{\rm ISR}}\phi(1020)f_0(980)$, and was subsequently
confirmed by BESII~\cite{y2175_bes}, Belle~\cite{y2175_belle} and
BESIII~\cite{y2175_bes3}. The observation of the $Y(2175)$ stimulated many
theoretical explanations of its nature, including a $s\bar{s}$-gluon
hybrid~\cite{hybrid}, an excited $\phi$ state~\cite{phiexcited}, a
tetraquark state~\cite{tetra} and a $\Lambda\bar{\Lambda}$ bound
state~\cite{bound}.  To study the $Y(2175)$, a dedicated data
set was collected with the BESIII detector~\cite{bes3}
at the BEPCII collider in 2015 at the center-of-mass energy ($\sqrt{s}$) of 2.125 GeV, which is in the vicinity of the peaking
cross sections for $e^+e^-\rightarrow\phi\pi\pi$ and $e^+e^-\rightarrow\phi f_0(980)$
decays reported by BaBar~\cite{babar_y2175_1, babar_y2175_2} and Belle~\cite{y2175_belle}.

In this paper, we present a determination of the integrated luminosity
of this data set using large-angle Bhabha scattering events
$e^+e^-\rightarrow (\gamma)e^+e^-$. A cross check is performed by
analyzing di-photon events $e^+e^-\rightarrow\gamma\gamma$.  In
addition, using the approach described in Ref.~\cite{ecms}, we
determine the center-of-mass energy using radiative dimuon events
$e^+e^-\rightarrow(\gamma)\mu^+\mu^-$, where
$\gamma$ represents possible ISR or FSR (final state radiation)
 photons.

\section{The BESIII detector}
BESIII~\cite{bes3} is a general purpose detector, which is located at
the BEPCII facility,
a double-ring $e^+e^-$ collider with a peak
luminosity of $10^{33}$~cm$^{-2}$s$^{-1}$ at a center-of-mass energy
of 3.773 GeV.  The BESIII detector covers 93\% of the solid angle
around the collision point and consists of four main components: 1) A
small-cell, helium-based main drift chamber (MDC) with 43 layers
providing an average single-hit resolution of 135 $\mu$m, and
charged-particle momentum resolution in a 1 T magnetic field of 0.5\%
at 1 GeV/$c$; 2) A Time-Of-Flight system (TOF) for
particle identification composed of a barrel and two end-caps. The
barrel has two layers, each consisting of 88 pieces of 5~cm
thick, 2.4~m long plastic scintillator.  Each end-cap consists of 96
fan-shaped, 5~cm thick, plastic scintillators.  The barrel (end-cap)
time resolution of 80~ps (110~ps) provides a $2\sigma$ $K/\pi$
separation for momenta up to about 1.0~GeV/$c$;
3) An electromagnetic calorimeter (EMC) consisting of
6240 CsI(Tl) crystals in a cylindrical structure, arranged in one barrel
and two end-caps.  The energy resolution for 1.0~GeV photons is 2.5\%
(5\%) in the barrel (end-caps), while the position resolution is 6~mm
(9~mm) in the barrel (end-caps);
4) A muon counter
(MUC) made of nine layers of resistive plate chambers in the barrel
and eight layers in each end-cap, which are incorporated in the iron return
yoke of the superconducting magnet.  The position resolution is about
2~cm.  A GEANT4~\cite{geant4_1, geant4_2}-based detector simulation
package has been developed to model the detector response.

\section{Monte Carlo simulation}
In order to determine the detection efficiency and estimate background
contributions, one million Monte Carlo (MC) events were simulated at
$\sqrt{s} = 2.125$~GeV for each of the four processes:
$e^+e^-\rightarrow(\gamma)e^+e^-$, $e^+e^-\rightarrow\gamma\gamma$,
$e^+e^-\rightarrow(\gamma)\mu^+\mu^-$ and $e^+e^-\rightarrow
q\bar{q}$.  The first three processes were generated with the Babayaga 3.5~\cite{babayaga} generator,
while $e^+e^-\rightarrow q\bar{q}\rightarrow \text{hadrons}$
was generated with EvtGen~\cite{besevtgen,evtgen} according to the
`LundAreaLaw'~\cite{lundmodel,lundarealaw}.

\section{Measurement of the luminosity}
\subsection{Event selection}
\label{sec:Bha_sel}
To select $e^+e^-\rightarrow(\gamma)e^+e^-$ events, exactly two good tracks with opposite charge were required. Each good charged
track was required to pass the interaction point within $\pm10$~cm
in the beam direction ($|V_z|<10.0$~cm) and within 1.0~cm in the plane
perpendicular to the beam ($V_r < 1.0$ cm). Their polar angles $\theta$ were required
to satisfy $|\cos\theta|<0.8$ to ensure the tracks were in the barrel
part of the detector.  The energy deposited in the EMC of each
track was required to be greater than $0.65\times E_{\text{beam}}$, where
$E_{\text{beam}} = 2.125/2$ GeV is the beam energy.  To select tracks
that were back-to-back in the MDC, $|\Delta\theta| \equiv |\theta_1 +
\theta_2 - 180^\circ| < 10^\circ$ and $|\Delta\phi| \equiv \left||\phi_1 -
\phi_2| - 180^\circ\right|<5.0^\circ$ were required, where $\theta_{1/2}$ and
$\phi_{1/2}$ are the polar and azimuthal angles of the two tracks,
respectively. Comparisons between data and MC simulation are shown in
Fig.~\ref{ee_sel}.

After applying the above requirements, 33,228,098 events were selected as Bhabha
scattering candidates. The background contribution is estimated to be
at the level of $10^{-5}$ using MC samples of
$e^+e^-\rightarrow\gamma\gamma$, $e^+e^-\rightarrow(\gamma)\mu^+\mu^-$
and $e^+e^-\rightarrow q\bar{q}$ processes, and is ignored in the calculation of
the integrated luminosity. The backgrounds from beam-gas interactions are also
ignored due to the powerful rejection rate of the trigger system and the distinguishable features of Bhabha events.

\subsection{Integrated luminosity}
\label{sec:Bha_lumi}
The integrated luminosity is calculated with
\begin{equation}
\label{lumi_formula}
L = \dfrac{N_{{\rm obs}}}{\sigma\times\varepsilon\times\varepsilon_{{\rm trig}}},
\end{equation}
where $N_{{\rm obs}}$ is the number of observed signal events, $\sigma$ is
the cross section of the specified process, $\varepsilon$ is the
detection efficiency and $\varepsilon_{{\rm trig}}$ is the trigger
efficiency.

For the Bhabha scattering process, the cross section at $\sqrt{s} = 2.125$~GeV is
calculated  with the Babayaga generator to be $1621.43 \pm 3.47$~nb.
Using the large sample of MC simulated events, the detection
efficiency is determined to be $(18.89\pm0.04)\%$.
The trigger efficiency $\varepsilon_{{\rm trig}}$ is 100\% with an
accuracy of better than 0.1\%~\cite{trigger}. The integrated
luminosity is determined to be $108.49\pm0.02\pm0.75$ pb$^{-1}$, where the first
uncertainty is statistical and the second one is systematic, which
will be discussed in Section~\ref{sec:Bha_syst}.

\end{multicols}
\ruleup
\begin{center}
\begin{minipage}{0.8\textwidth}
\centering
\includegraphics[width=0.48\textwidth]{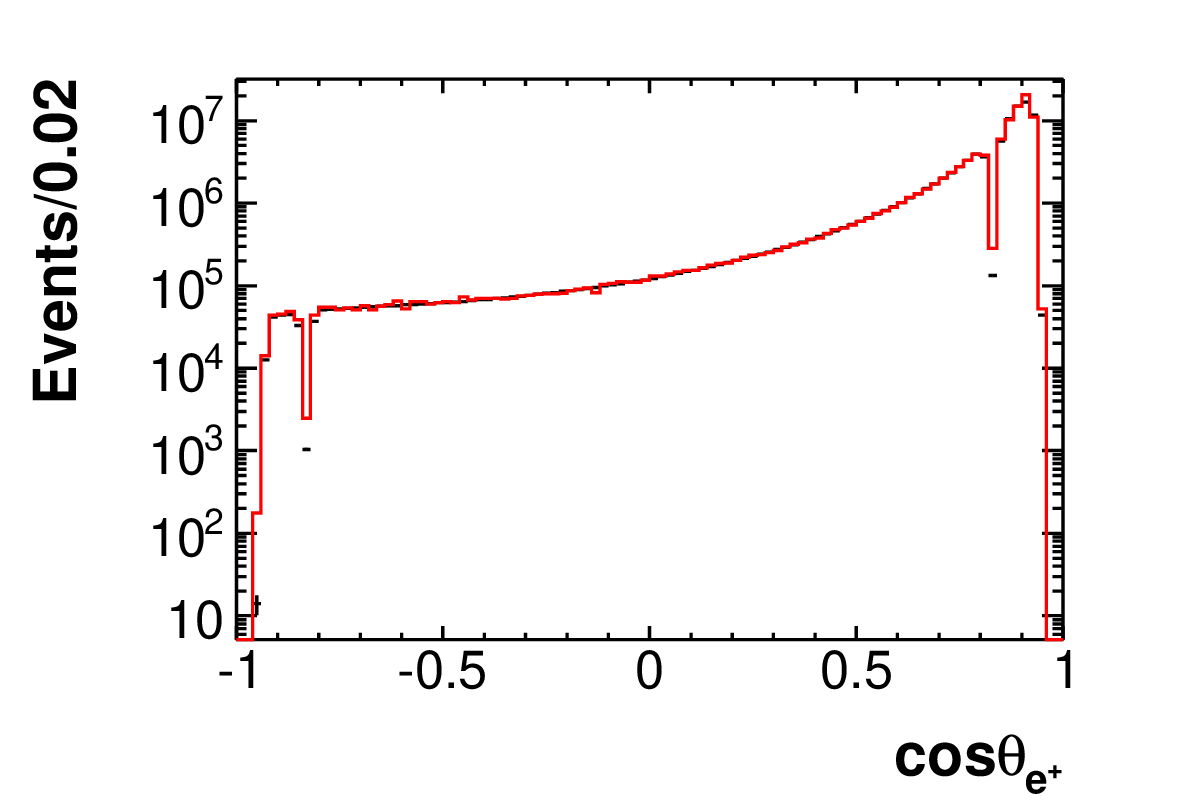}
\put(-120,100){(a)}
\includegraphics[width=0.48\textwidth]{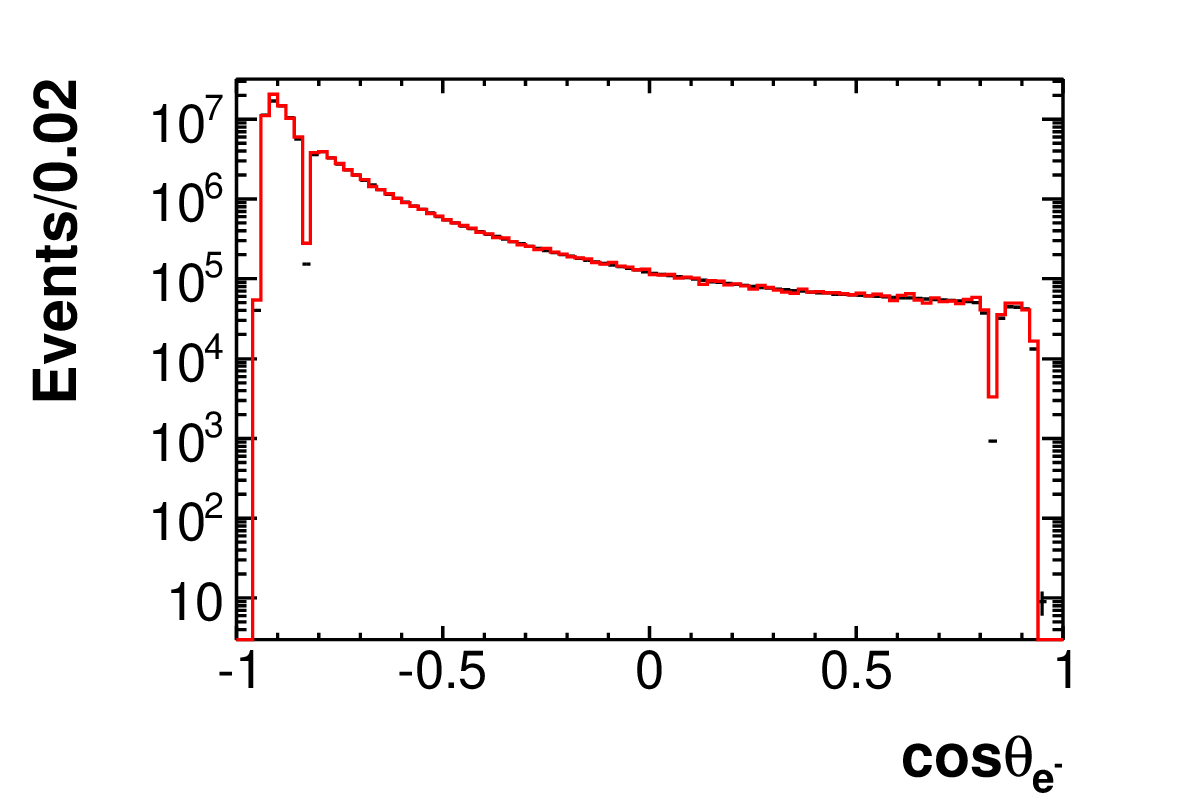}
\put(-120,100){(b)}

\includegraphics[width=0.48\textwidth]{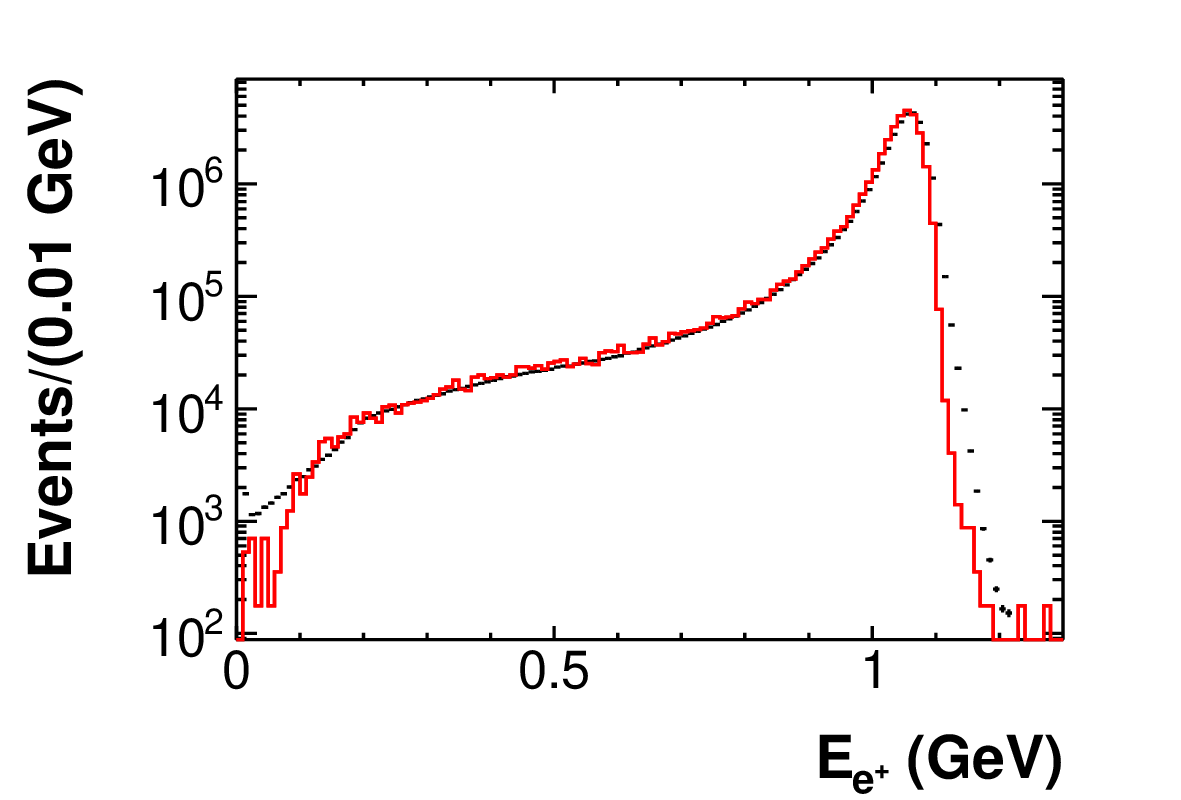}
\put(-120,100){(c)}
\includegraphics[width=0.48\textwidth]{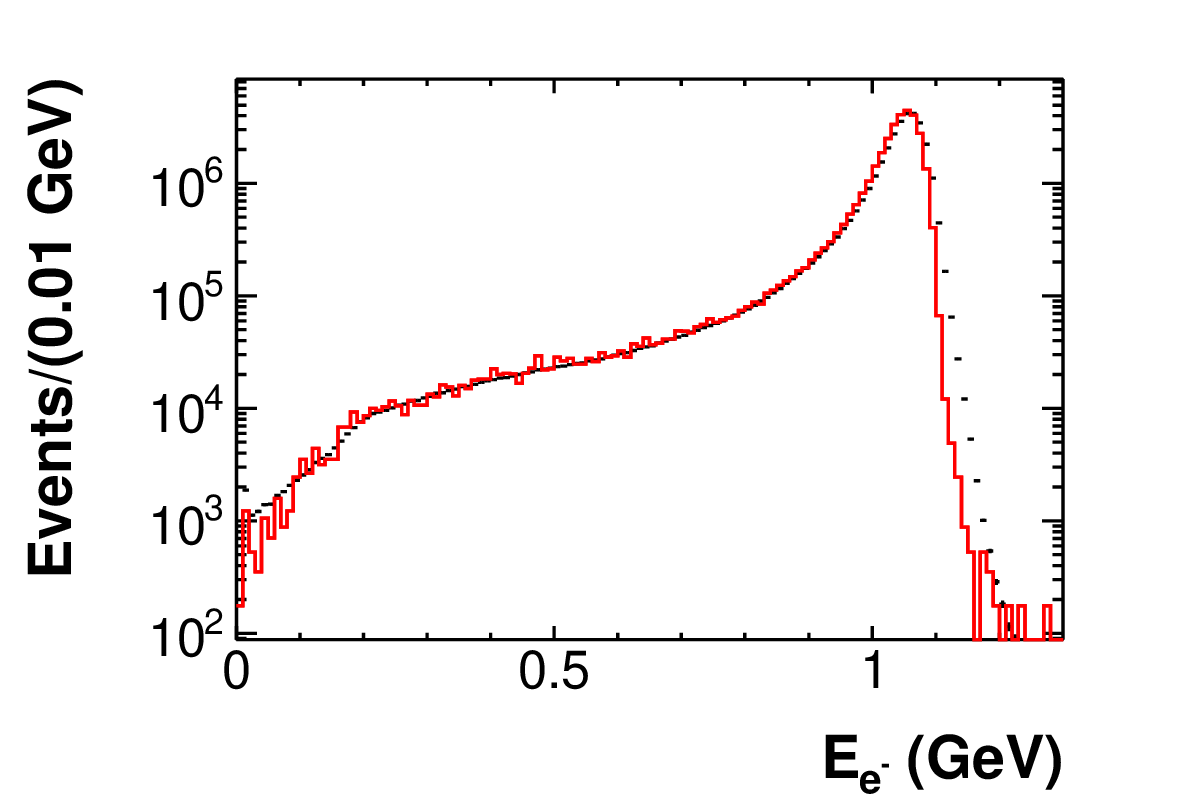}
\put(-120,100){(d)}

\includegraphics[width=0.48\textwidth]{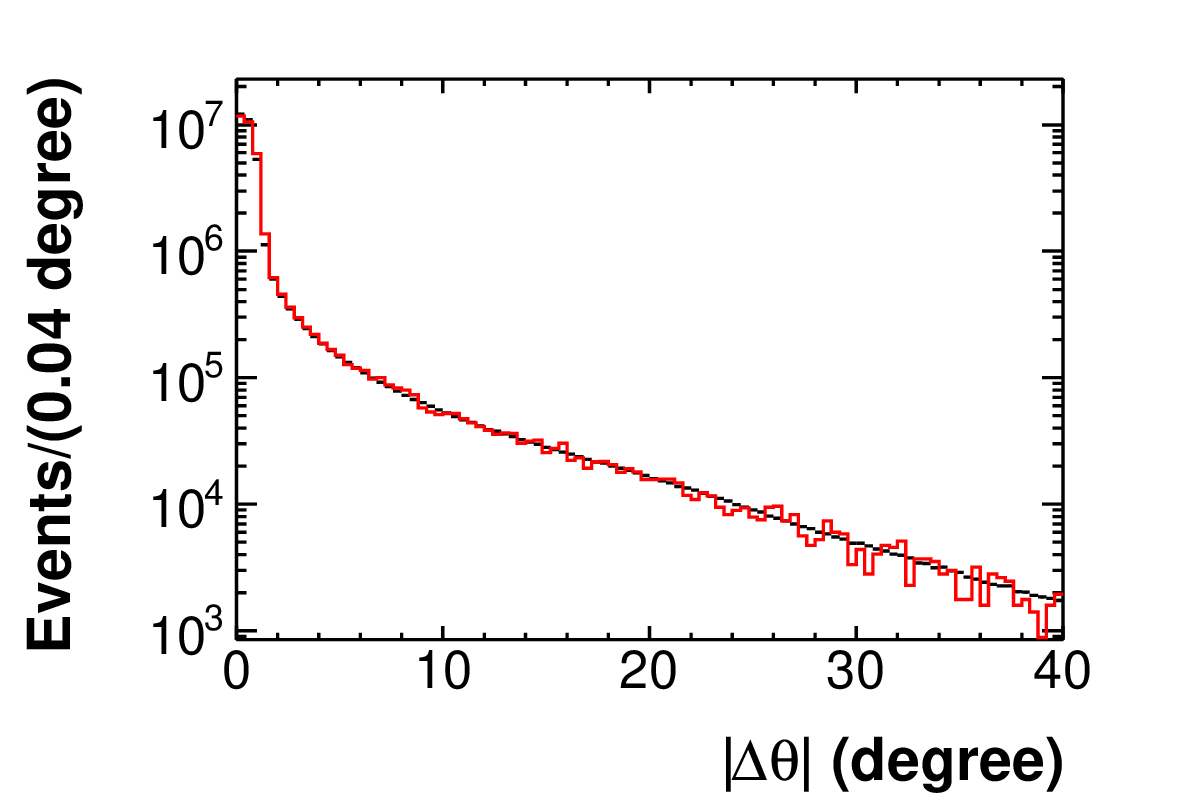}
\put(-120,100){(e)}
\includegraphics[width=0.48\textwidth]{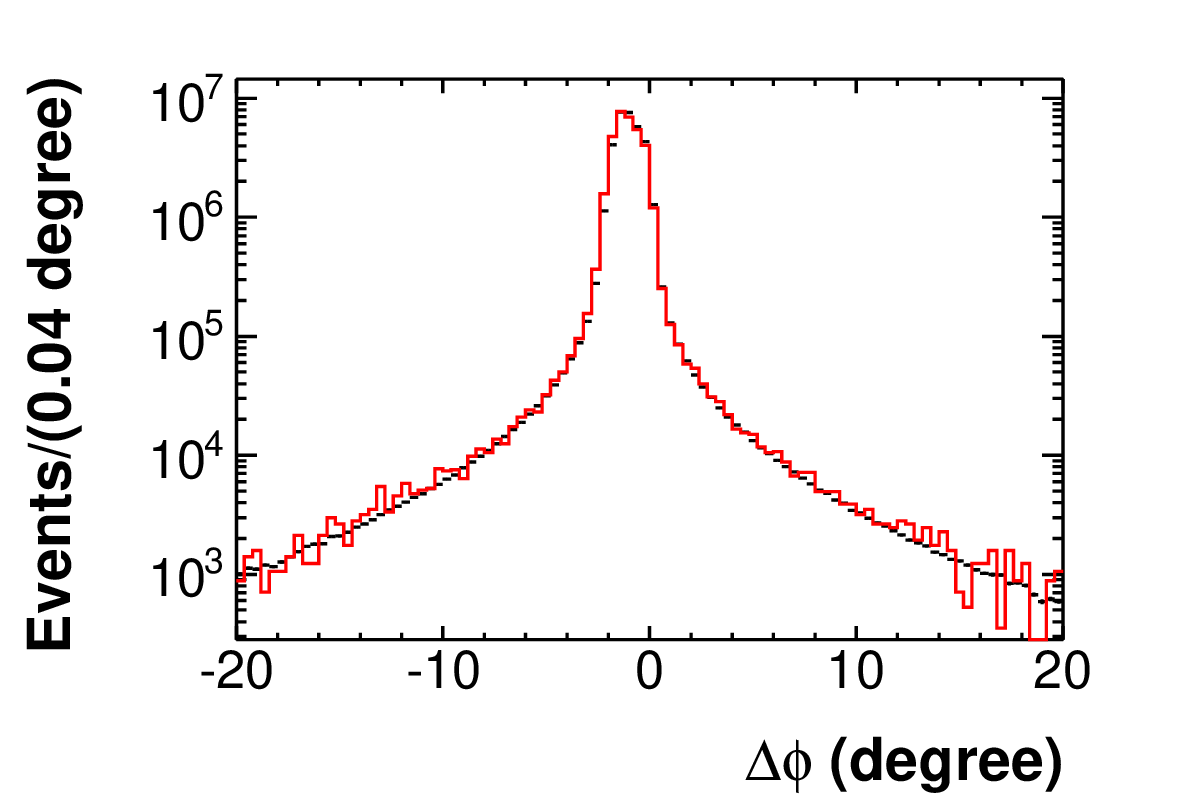}
\put(-120,100){(f)} \figcaption{\label{ee_sel} Distributions of $\cos\theta$
  of (a) $e^+$ and (b) $e^-$, deposited energy in the EMC
  of (c) $e^+$ and (d) $e^-$, (e) $|\Delta\theta|$
   and
  (f) $\Delta\phi$ (measured in the laboratory frame of reference).
  The dots with error bars are for data, while the solid line
  indicates signal MC simulation.}
 \end{minipage}
 \end{center}
\ruledown
\begin{multicols}{2}

\subsection{Systematic uncertainty}
\label{sec:Bha_syst}
Sources of systematic uncertainty include the requirements on track angles ($\theta$,
 $\Delta\theta$, $\Delta\phi$) and the deposited energy in the EMC,
the tracking efficiency, beam energy, MC statistics, trigger
efficiency, and the MC generator.

To estimate the systematic uncertainties associated with
the related angular requirements, the same selection criteria with alternative
quantities were performed, individually, and the resultant (largest) difference with respect to
the nominal result taken as the systematic uncertainty:
 $|\cos\theta|< 0.8$ was changed to $|\cos\theta|
< 0.75$, resulting in a relative difference to the nominal result of 0.06\%;
 $|\Delta\theta|<10.0^\circ$ was
changed to $8.0^\circ$ or $15.0^\circ$, and the systematic uncertainty
estimated to be 0.02\%; $|\Delta\phi| < 5.0^\circ$ was changed to
$4.0^\circ$ or $10.0^\circ$, and the associated systematic uncertainty is
0.04\%.

The uncertainty associated with the requirement on the deposited energy in the EMC
is determined by comparing the detection efficiency between data and
MC simulation. The data and MC samples were selected using the
selection criteria listed in Section~\ref{sec:Bha_sel}
except for the deposited energy requirement on the electron/positron.
The efficiency
is determined by the ratio between the numbers of events with and
without the deposited energy requirement. The difference in the
detection efficiency between data and signal MC
simulation is 0.19\% and 0.13\% for electrons and positrons, respectively.
The sum, 0.32\%, is taken as the
 systematic uncertainty.

For the uncertainty associated with the tracking efficiency, it has been well studied in Ref.~\cite{rscan_lumi} by selecting
a control sample of Bhabha events with the EMC information only. It was found that the difference between data and MC
simulation is 0.41\%, which is taken as the systematic uncertainty.

To estimate the systematic uncertainty associated with the beam energy, the
luminosity is recalculated with the updated cross section and
detection efficiency at the alternative center-of-mass energy of the measured value in Section~\ref{sec:cm_measurement}. The difference
from the nominal luminosity, 0.18\%, is taken as the systematic
uncertainty.

The uncertainty from MC statistics is 0.21\% and from trigger
efficiency is 0.1\%~\cite{trigger}. The uncertainty due to the
Babayaga generator is given as 0.5\%~\cite{babayaga}.

All individual systematic uncertainties are summarized in Table~\ref{tab_syst}.
Assuming the individual uncertainties to be independent, the total
systematic uncertainty is calculated by adding them quadratically and found to
be 0.78\%.

\begin{center}
\tabcaption{\label{tab_syst} Summary of the systematic uncertainties.}
\footnotesize
\begin{tabular*}{80mm}{c@{\extracolsep{\fill}}c}
\toprule Source & Relative uncertainty (\%) \\
\hline
$|\cos\theta| < 0.8$ & 0.06 \\
$|\Delta\theta| < 10.0^\circ$ & 0.02 \\
$|\Delta\phi| < 5.0^\circ$ & 0.04 \\
Deposited energy requirement & 0.32 \\
Tracking efficiency & 0.41 \\
Beam energy & 0.18 \\
MC statistics & 0.21 \\
Trigger efficiency & 0.10 \\
Generator & 0.50 \\
\hline
Total & 0.78 \\
\bottomrule
\end{tabular*}
\vspace{0mm}
\end{center}

\subsection{Cross check}
\label{sec:cross_check}
As a cross check, an alternative luminosity measurement using $e^+e^-\rightarrow\gamma\gamma$ events was
performed. To select $e^+e^-\rightarrow\gamma\gamma$ events, candidate
events must have two energetic clusters in the EMC. For each cluster,
the polar angle was required to satisfy $|\cos\theta| < 0.8$ and the deposited
energy $E$ must be in region $0.7\times E_{\text{beam}} < E < 1.15\times E_{\text{beam}}$. To select clusters that are
back-to-back, $|\Delta\phi|< 2.5^\circ$ (defined in
Section~\ref{sec:Bha_sel}) was required.  In addition, there should be
no good charged tracks satisfying $|V_z| < 10.0$~cm and $V_r <
1.0$~cm. With the selected $e^+e^-\rightarrow\gamma\gamma$ events, the integrated
luminosity is determined to be $107.91\pm0.05$~pb$^{-1}$ (statistical only), which is in
good agreement with the result obtained using large-angle Bhabha scattering events.

\section{Measurement of the center-of-mass energy}
\label{sec:cm_measurement}

\subsection{Event selection}
To select $e^+e^-\rightarrow(\gamma)\mu^+\mu^-$ candidates, we
require exactly two good tracks with opposite charge satisfying
$|V_z|<10.0$~cm, $V_r<1.0$~cm and $|\cos\theta|<0.8$. To remove Bhabha
events, the ratio of the deposited energy in the
EMC and the momentum of a charged track, $E/pc$, was required to be less
than 0.4. The two tracks should be back-to-back,
with the $\Delta\theta$ and $\Delta\phi$ (defined in Section~\ref{sec:Bha_sel}) satisfying
 $|\Delta\theta|
< 10.0^\circ$ and $|\Delta\phi| < 5.0^\circ$.
To further suppress background from cosmic rays, $|\Delta
T| = |t_1 - t_2| < 1.5$~ns was required, where $t_{1/2}$ is the time of flight of the two charged
tracks recorded by the TOF. Figure~\ref{uu_sel} shows the comparisons
between data and MC simulation, where the solid line is signal MC
and the shaded histogram represents the simulation of background $e^+e^-\rightarrow
q\bar{q}$.

With the above requirements, 1,472,195 events were selected in
data with an estimated background level of about 1.8\%. The small
bumps visible in Figs.~\ref{uu_sel} (g) and (h) at about 0.93
GeV/$c$ mainly come from the $e^+e^-\rightarrow K^+K^-$ process.
The peak at about 1.07 GeV/$c$ mainly consists of events from the processes
$e^+e^-\rightarrow\pi^+\pi^-$ and $e^+e^-\rightarrow\pi^+\pi^-\gamma$.

\end{multicols}
\ruleup
\begin{center}
\begin{minipage}{1\textwidth}
\centering
\includegraphics[width=0.39\textwidth]{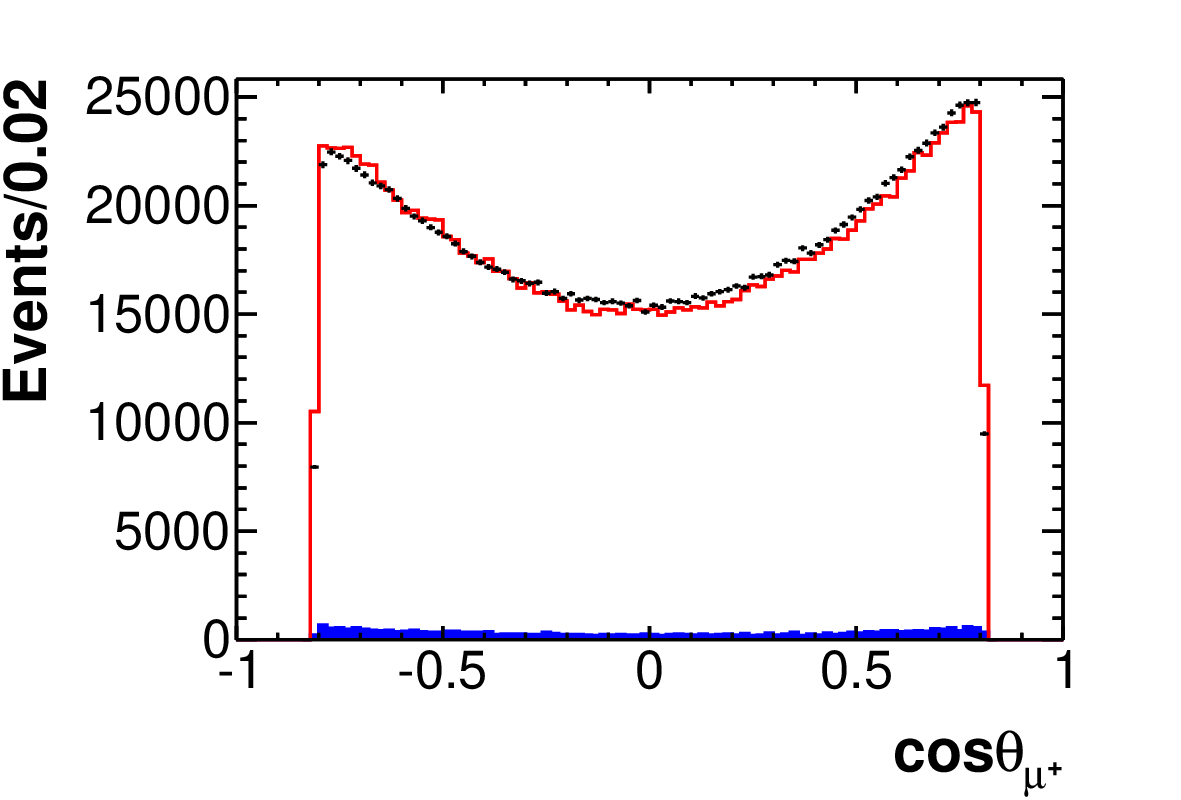}
\put(-110, 100){(a)}
\includegraphics[width=0.39\textwidth]{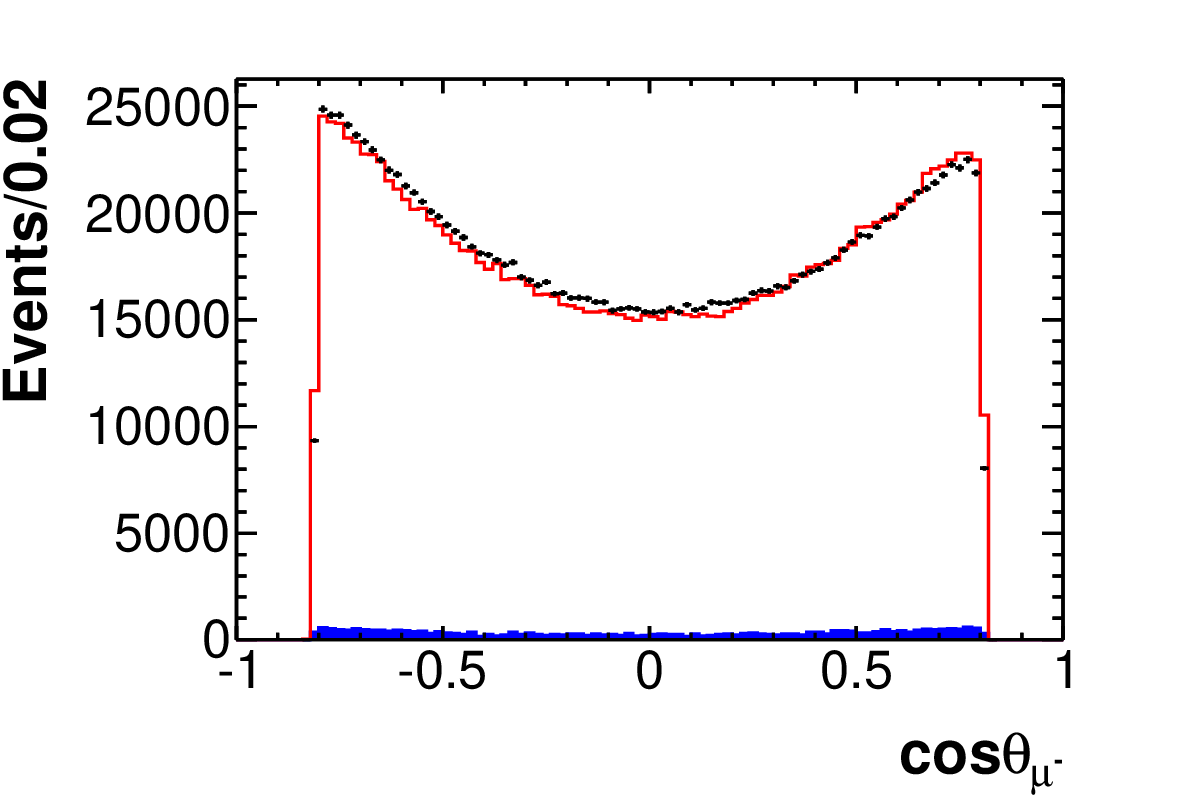}
\put(-110, 100){(b)}

\includegraphics[width=0.39\textwidth]{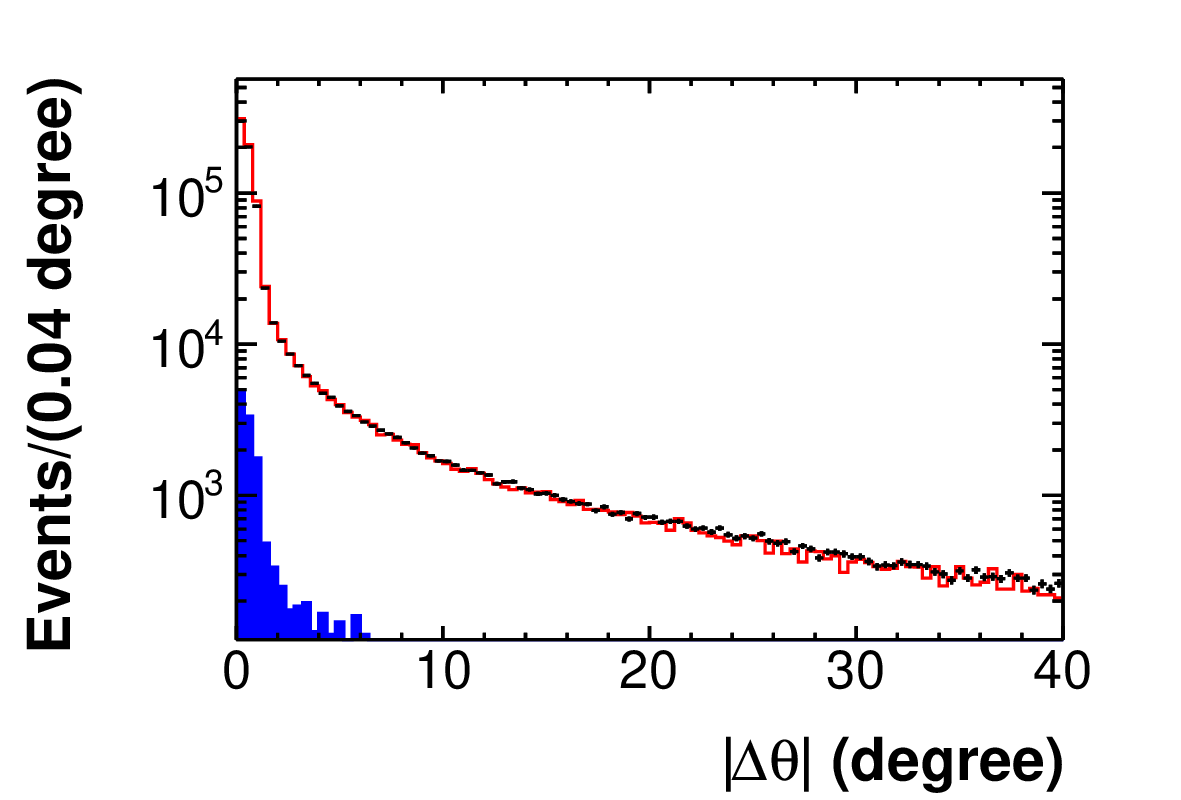}
\put(-140, 100){(c)}
\includegraphics[width=0.39\textwidth]{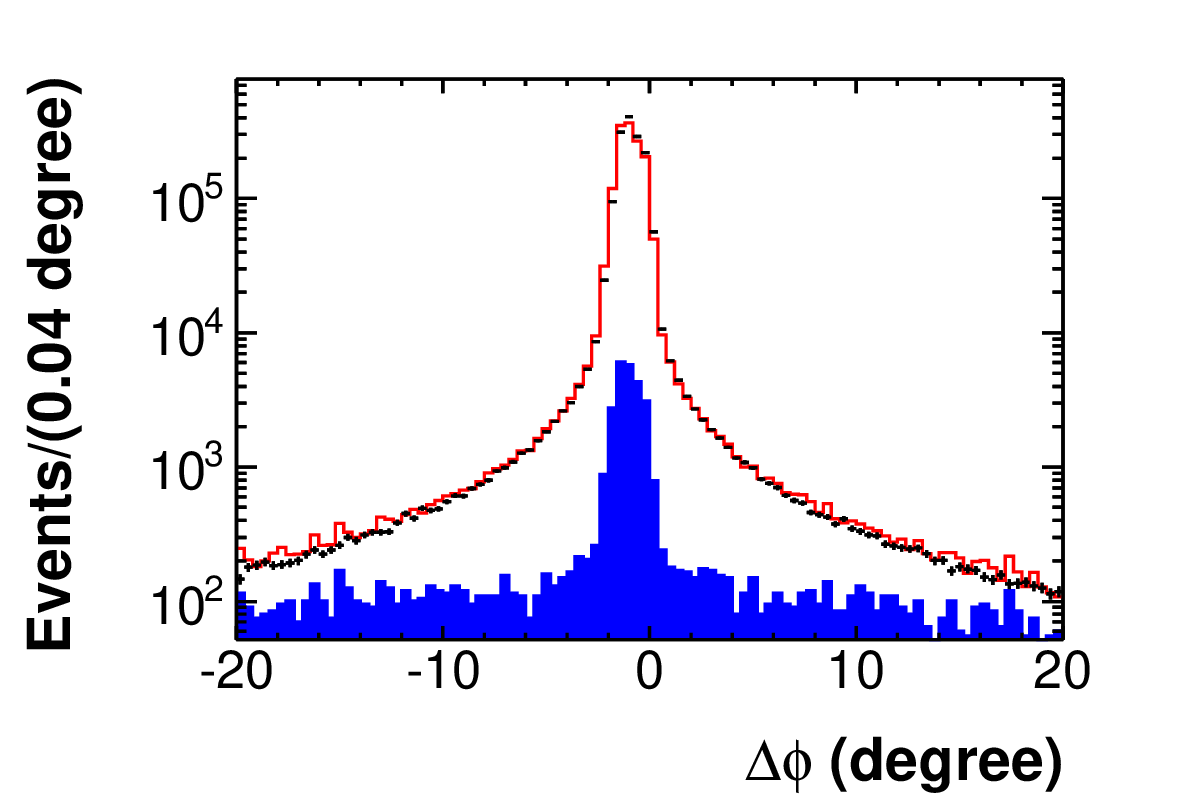}
\put(-140, 100){(d)}

\includegraphics[width=0.39\textwidth]{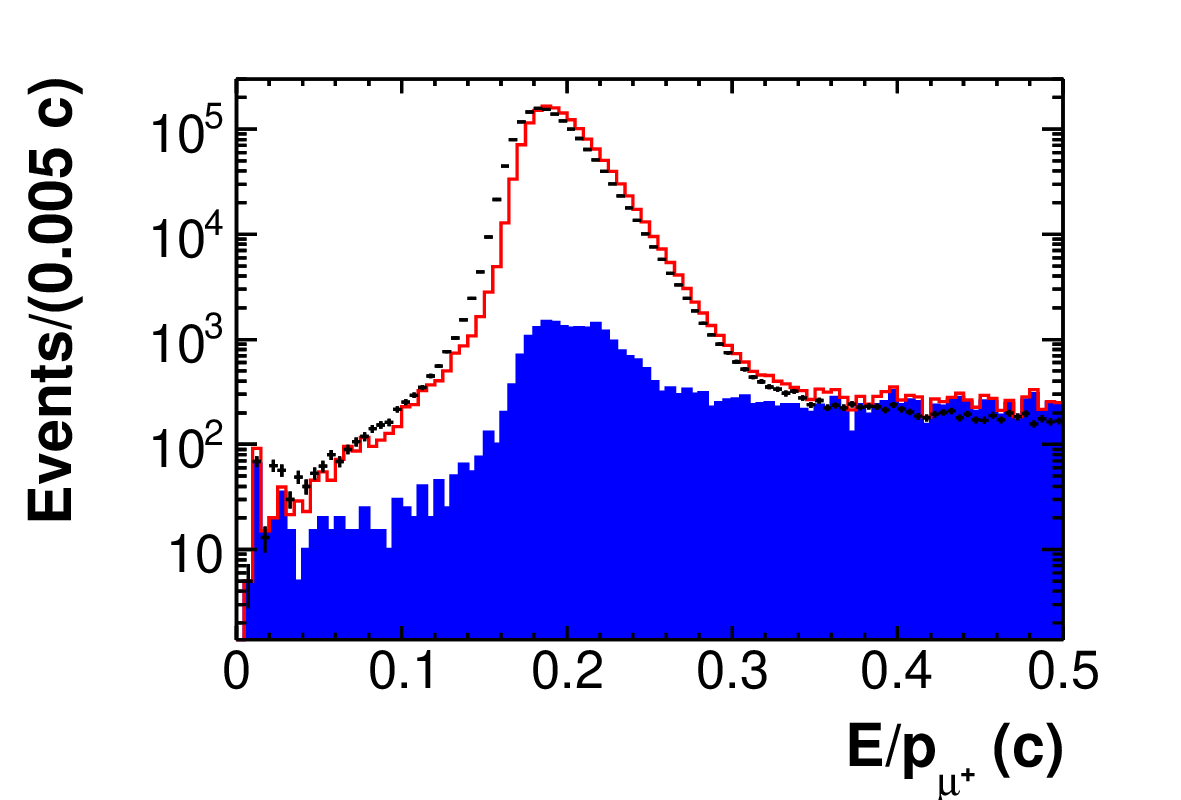}
\put(-140, 100){(e)}
\includegraphics[width=0.39\textwidth]{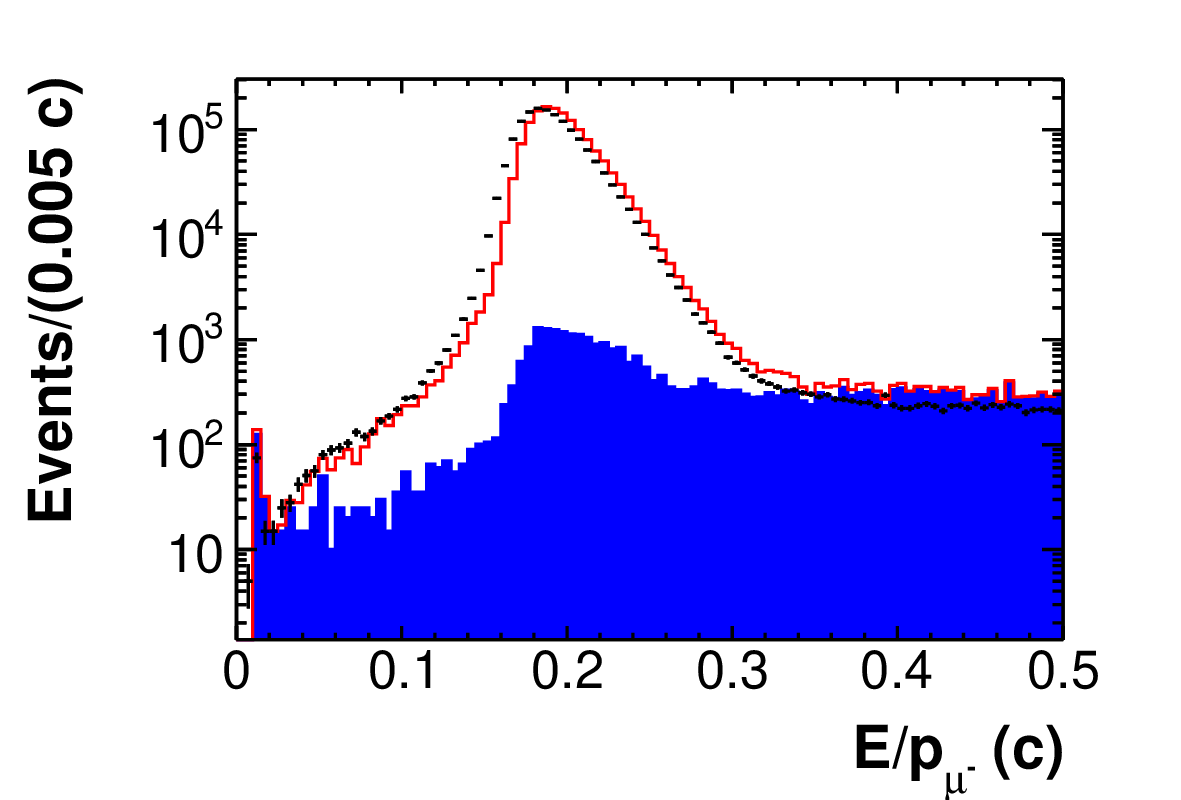}
\put(-140, 100){(f)}

\includegraphics[width=0.39\textwidth]{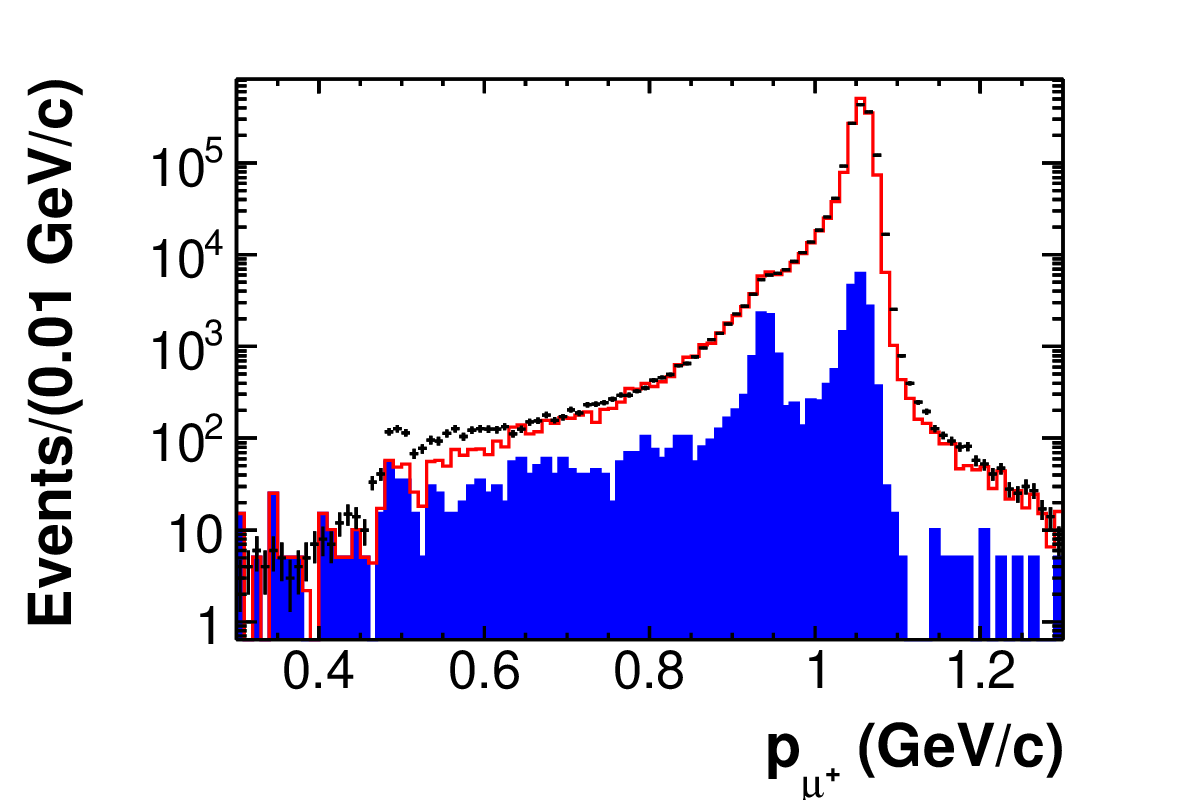}
\put(-140, 100){(g)}
\includegraphics[width=0.39\textwidth]{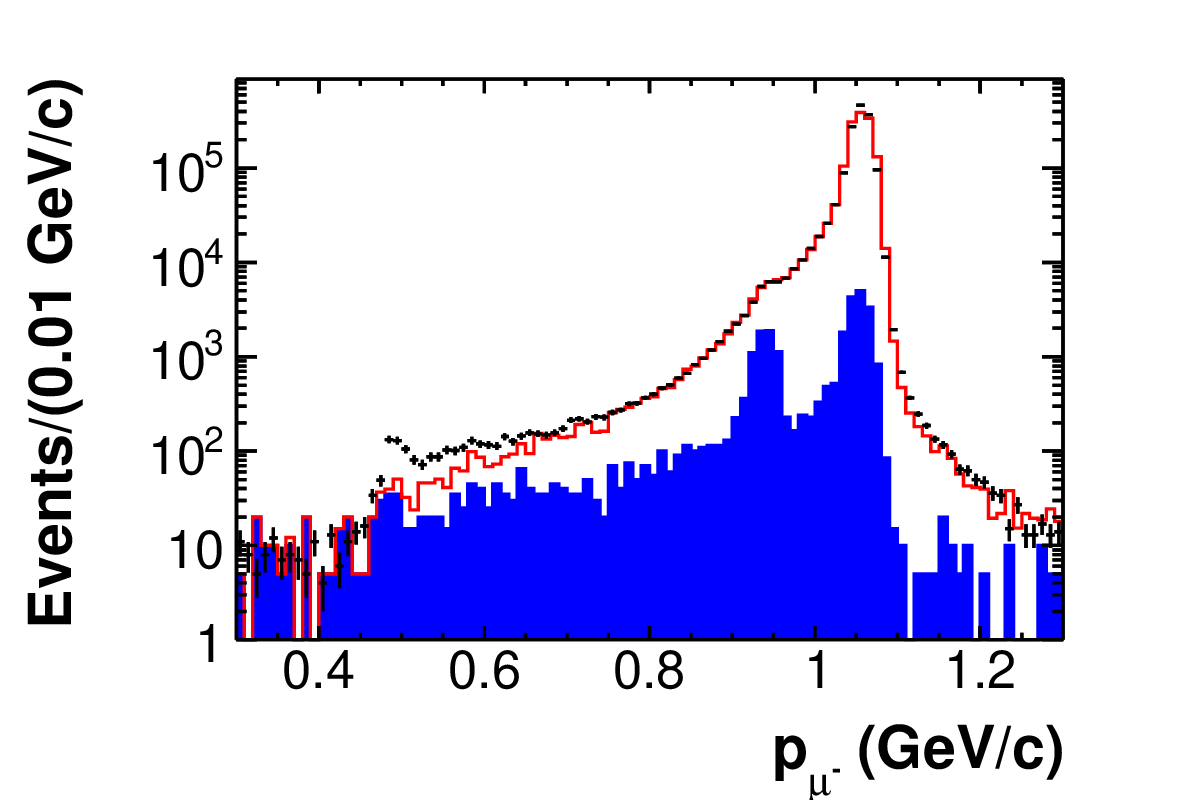}
\put(-140, 100){(h)}

\includegraphics[width=0.39\textwidth]{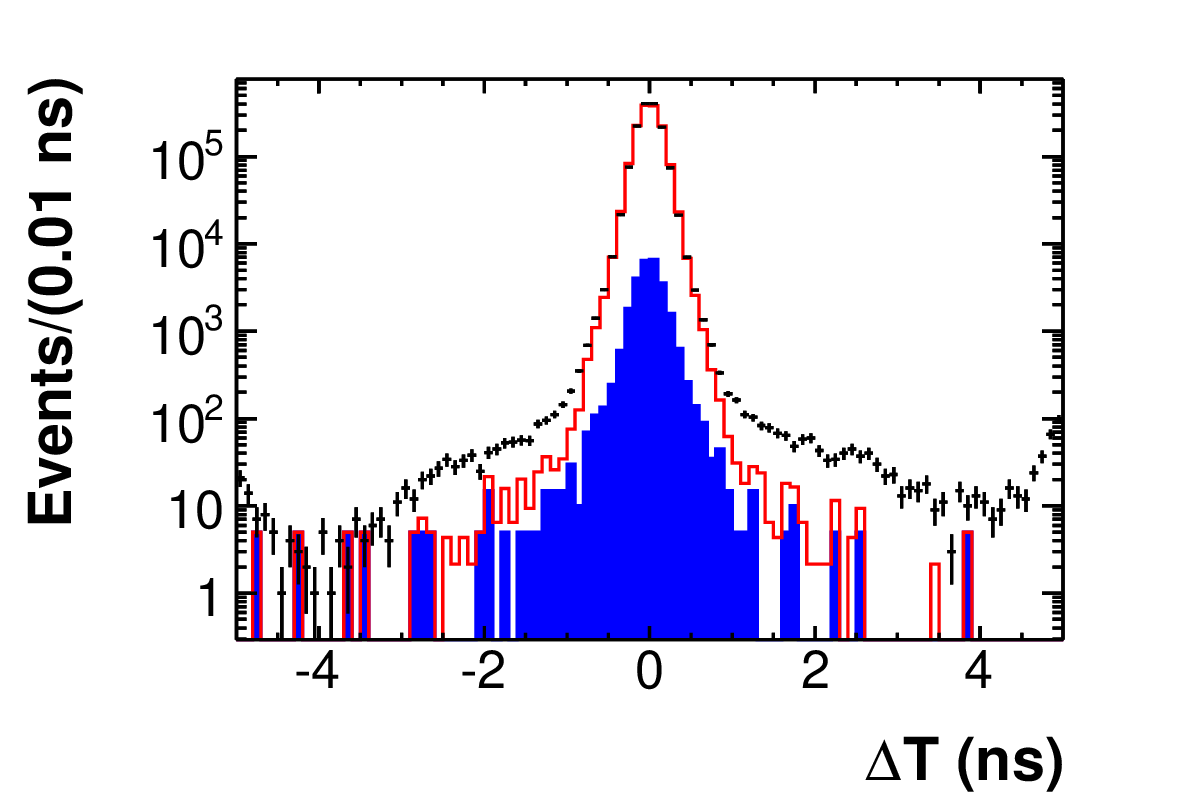}
\put(-140, 100){(i)}
 \figcaption{\label{uu_sel} Distributions of $\cos\theta$
  of (a) $\mu^+$ and (b) $\mu^-$, (c) $|\Delta\theta|$,
  and (d) $\Delta\phi$ (measured in the laboratory frame of reference), $E/p$ distributions of
  (e) $\mu^+$ and (f) $\mu^-$,
  momentum distributions of (g) $\mu^+$ and (h) $\mu^-$, and (i) $\Delta T$ distribution.
  The dots with error bars represent the data,
   the solid line indicates signal MC simulation, and the shaded
  histogram represents the MC simulation of $e^+e^-\rightarrow q\bar{q}$.}
\end{minipage}
\end{center}
\ruledown

\begin{multicols}{2}
\subsection{Center-of-mass energy}
\label{sec:cms energy}
Using the $e^+e^-\rightarrow(\gamma)\mu^+\mu^-$ events,
the center-of-mass energy of the data set is determined
with the method described in Ref.~\cite{ecms}.

The center-of-mass energy can be determined with
\begin{equation}
\label{mcm}
M_{\text{CM}} = M_{\text{data}}(\mu^+\mu^-) - \Delta M,
\end{equation}
where $M_{\text{data}}(\mu^+\mu^-)$ is the reconstructed $\mu^+\mu^-$ invariant mass of
the selected $e^+e^-\rightarrow(\gamma)\mu^+\mu^-$ events, and
$\Delta M$ is the correction for effects of ISR and FSR, which can be estimated using
the $\mu^+\mu^-$ invariant mass of MC
samples with ISR/FSR turned on ($M_{\text{MC, on}}(\mu^+\mu^-)$) and off ($M_{\text{MC, off}}(\mu^+\mu^-)$):
\begin{equation}
\label{delta_isrfsr}
\Delta M = M_{\text{MC, on}}(\mu^+\mu^-) - M_{\text{MC, off}}(\mu^+\mu^-).
\end{equation}

By fitting the $M_{\text{MC, on}}(\mu^+\mu^-)$ and $M_{\text{MC, off}}(\mu^+\mu^-)$  distributions of MC samples, the average of $\Delta M$
is determined to be $-1.13$~MeV/$c^2$, where $M_{\text{MC, on}}=2124.60\pm0.04$~MeV/$c^2$, $M_{\text{MC, off}}=2125.73\pm0.01$~MeV/$c^2$, and the errors are
statistical only.
The function fitted to $M_{\text{MC, on}}$ is a Gaussian plus a Crystal Ball
function~\cite{crystal_ball} with a common mean, and the function fitted to $M_{\text{MC, off}}$ is a double Gaussian function with a common mean.
The fit results are shown in Fig.~\ref{muu_uu}.
To calculate $\Delta M$ (and $M_{{\rm CM}}$) as a function of run number, $M_{\text{MC, off}}$
and $M_{\text{MC, on}}$ for each run are fitted with the above same functions.

\end{multicols}
\ruleup
\begin{center}
\begin{minipage}{1.0\textwidth}
\includegraphics[width=0.48\textwidth]{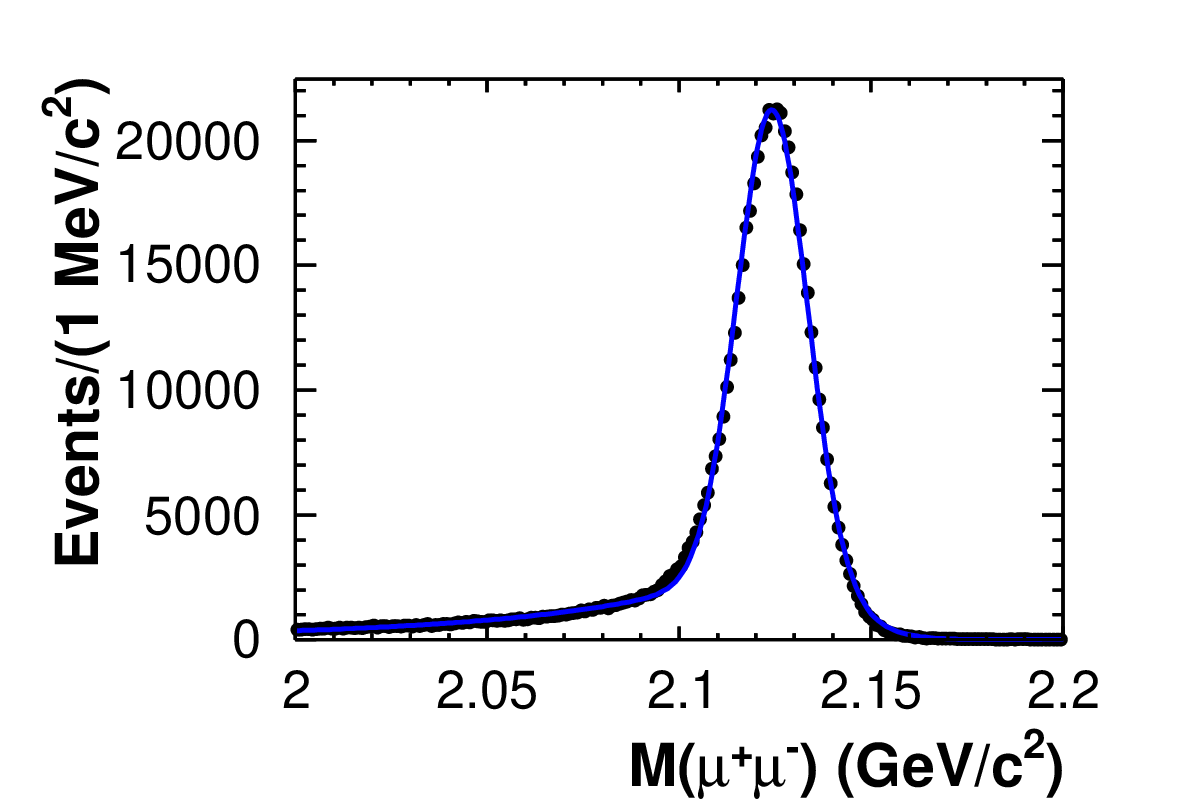}
\put(-130,110) {(a)}
\hfill
\includegraphics[width=0.48\textwidth]{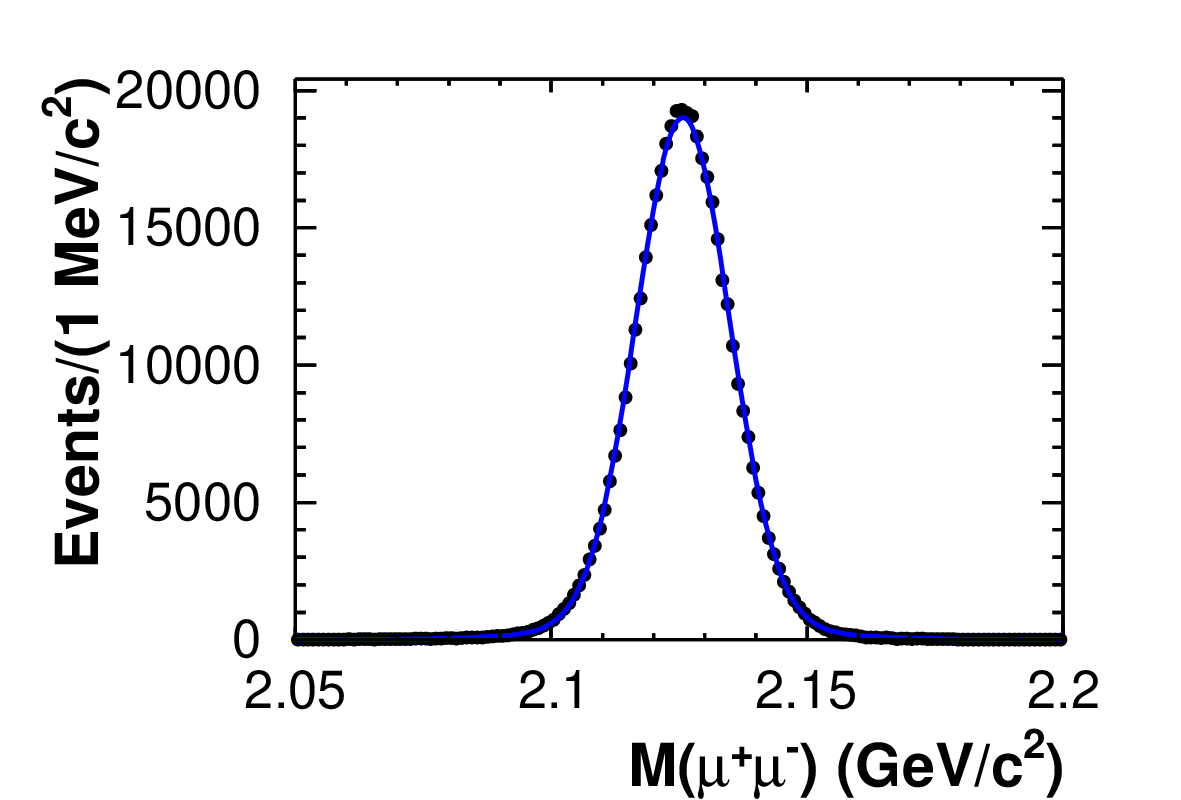}
\put(-130,110) {(b)}
\figcaption{\label{muu_uu} Fit to $M(\mu^+\mu^-)$ of MC sample (a) with and (b) without ISR/FSR.}
\vspace{10pt}
\end{minipage}
\end{center}
\ruledown
\begin{multicols}{2}

For each run, $M_{\text{data}}(\mu^+\mu^-)$ was fitted in the range [2.0, 2.2]~GeV/c$^2$.  The signal is described by a Gaussian plus a Crystal
Ball function with a common mean, while the background is ignored
(about 1.1\% in the fit range).  As an example, the fit result for run
42030 is shown in Fig.~\ref{muu_data}.
\begin{center}
\begin{minipage}{0.5\textwidth}
\includegraphics[width=1\textwidth]{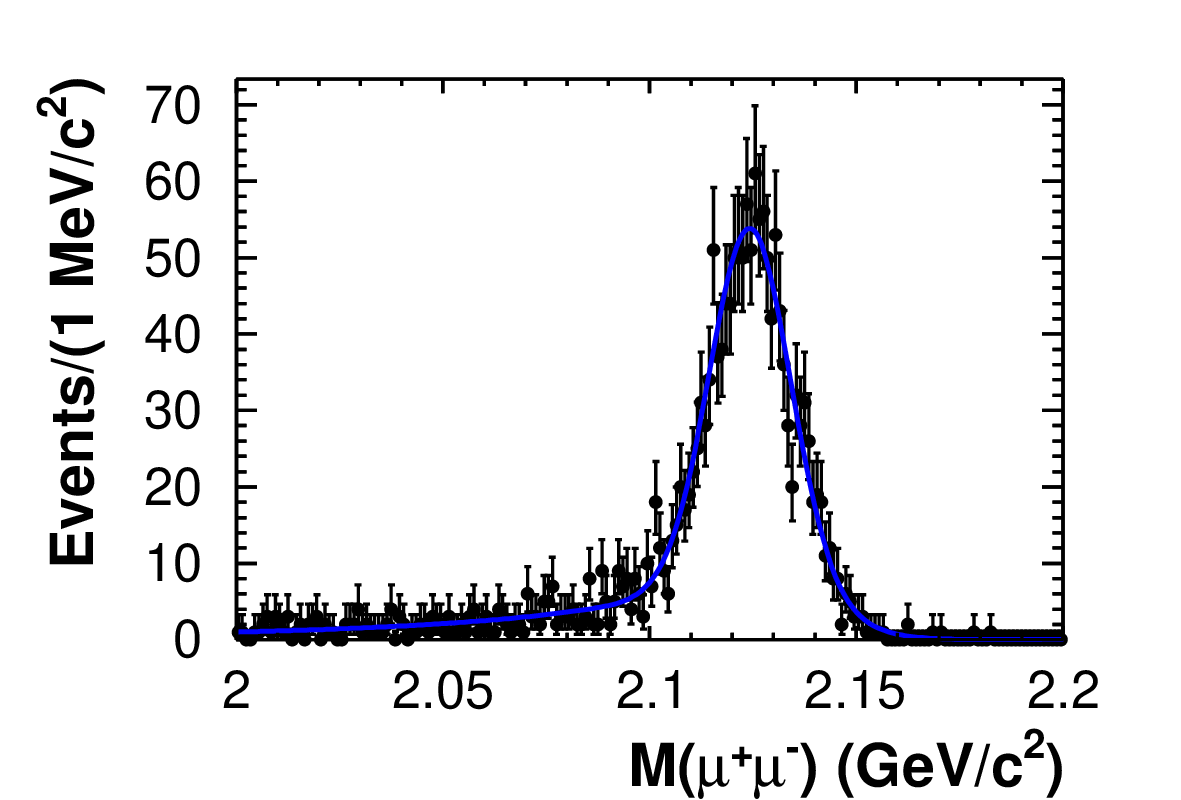}
\figcaption{\label{muu_data} Fit to $M(\mu^+\mu^-)$ for run
  42030. Dots with error bars are data, while the solid line is the fit result.}
\end{minipage}
\end{center}

\begin{center}
\begin{figure}[!htbp]
\includegraphics[width=1\textwidth]{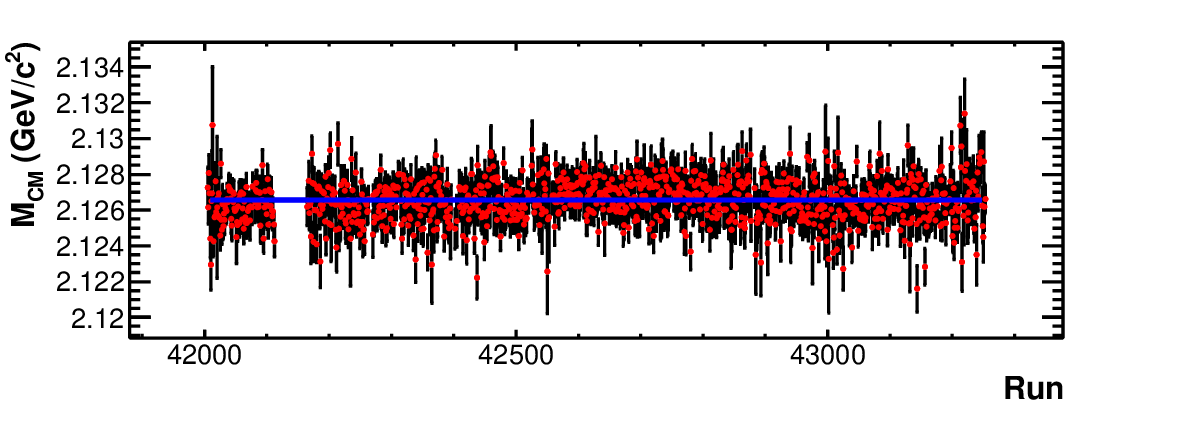}
\figcaption{\label{mcm_good} Distribution of $M_{\text{CM}}$ for individual runs. The solid line is the average fit.}
\end{figure}
\end{center}

$M_{\text{CM}}$ was calculated with Eq.~\ref{mcm} and
Eq.~\ref{delta_isrfsr} for each run.  The average of $M_{\text{CM}}$
for the full data set is determined to be $2126.55\pm0.03$
MeV/$c^2$ by fitting the $M_{\text{CM}}$ of different runs with a
constant. The $M_{\text{CM}}$ distribution as a function of run number
and the overall fit result
are shown in Fig.~\ref{mcm_good}, where 21 runs are excluded in the
fit due to large statistical errors (less than 100 entries in the
fit range).
The $M_{\text{CM}}$ values for individual runs are shown in Fig.~\ref{hist_mcm}
as a histogram, which can be fitted very well with a Gaussian function with the parameters
$\mu=2126.55\pm0.03$ MeV/$c^2$ and $\sigma = 0.98\pm0.03$ MeV/$c^2$.

\end{multicols}
\ruleup
\ruledown
\begin{multicols}{2}

\begin{center}
\begin{minipage}{0.5\textwidth}
\includegraphics[width=1\textwidth]{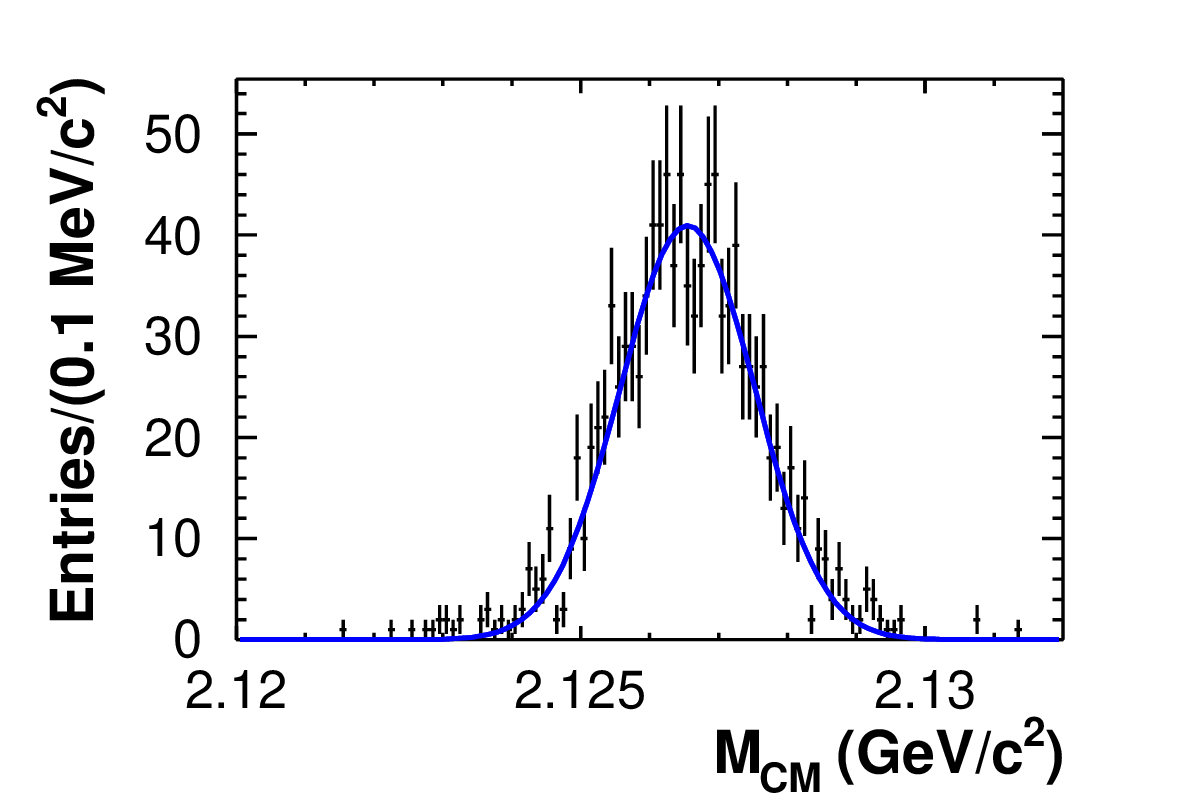}
\figcaption{\label{hist_mcm} Histogram of $M_{\text{CM}}$ for individual runs. The solid line is a Gaussian function.}
\end{minipage}
\end{center}

\subsection{Systematic uncertainty}

As shown in Section~\ref{sec:cms energy},
$M_{\text{MC, off}} = 2125.73\pm0.01$~MeV/c$^2$
is 0.73~MeV/c$^2$ higher than the input value (2125~MeV/c$^2$). This
difference is taken as the systematic uncertainty.

The uncertainty of the momentum measurement of two muon tracks has been studied using
$e^+e^-\rightarrow\gamma_{{\rm ISR}}J/\psi, J/\psi\rightarrow\mu^+\mu^-$
in Ref.~\cite{ecms}, and is estimated to be 0.011\%.

To estimate the uncertainty from the fit to the invariant mass of $\mu^+\mu^-$,
the signal shape and fit range were varied and $M_{{\rm CM}}$ re-calculated.
The difference to the nominal result is taken as the systematic uncertainty.
The systematic uncertainty from signal shape is estimated to be 0.08 MeV/c$^2$ by
replacing the Crystal Ball function in the signal shape with the GaussExp function~\cite{gaussexp}.
The systematic uncertainty from fit range is estimated to be 0.13 MeV/c$^2$ by
varying the fit range to [2.00, 2.14] GeV/c$^2$ and [2.10, 2.20] GeV/c$^2$, respectively.

To estimate the uncertainty from the fit to the
$M_{{\rm CM}}$ distribution, fits in different ranges of
the run number were carried out.
The resultant maximum difference with respect to the nominal value,
0.34~MeV/c$^2$, is taken as the uncertainty.

Assuming all of the above uncertainties are independent,
the total systematic uncertainty is calculated to
be 0.85~MeV/c$^2$ by adding the individual items in quadrature.

\section{Summary}
The integrated luminosity of the data taken at 2.125~GeV in 2015 with the BESIII
detector is
measured to be $108.49\pm0.02\pm0.85$~pb$^{-1}$ using large-angle
Bhabha events.  A cross check with $e^+e^-\rightarrow\gamma\gamma$
events was performed and the result is $107.91\pm0.05$~pb$^{-1}$ (statistical only), which
is in good agreement with the nominal result within the uncertainties.  With
$e^+e^-\rightarrow(\gamma)\mu^+\mu^-$ events, the center-of-mass
energy of the data set is measured to be $2126.55\pm0.03\pm0.85$ MeV.
The results in this measurement are important input for physics studies,
{\it e.g.}, studies of decays of the $Y(2175)$.
\\

\acknowledgments{The BESIII collaboration would like to thank the staff of
BEPCII and the IHEP computing center for their dedicated support.}
\end{multicols}

\vspace{-1mm}
\centerline{\rule{80mm}{0.1pt}}
\vspace{2mm}

\begin{multicols}{2}

\end{multicols}

\clearpage
\end{CJK*}
\end{document}